\documentclass[fleqn,aps,pre,preprint,a4paper,onecolumn]{revtex4}

\usepackage[]{amsmath}
\usepackage{amssymb}
\usepackage{graphicx}
\usepackage{color}
\usepackage{tabularx}
\usepackage{mathtools}
\usepackage{algpseudocode}
\usepackage{enumitem}
\usepackage{multirow}
\usepackage{epstopdf}  
\usepackage{yfonts}


\newcommand{\vect}[1]{\textbf{\textit{#1}}}

\newcommand{\limit}{{\mathrm{limit}}}
\newcommand{\start}{{\mathrm{start}}}
\newcommand{\olayer}{{\mathrm{out}}}

\newcommand{\mr}{\mathcal {R}}
\newcommand{\ml}{\mathcal {L}}
\newcommand{\mn}{\mathcal {N}}

\newcommand{\neighbor}{{ {N}}}

\newcommand {\newparagraph} {\vskip .3cm\noindent}

\begin{document}

\title{DeePMD-kit: A deep learning package for many-body potential energy representation
and molecular dynamics}
\author{Han Wang}
\email{wang_han@iapcm.ac.cn}
\affiliation{Institute of Applied Physics and Computational Mathematics,
Fenghao East Road 2, Beijing 100094, P.R.~China}
\affiliation{CAEP Software Center for High Performance Numerical
Simulation, Huayuan Road 6, Beijing 100088, P.R.~China}
\author{Linfeng Zhang}
\email{linfengz@princeton.edu}
\affiliation{Program in Applied and Computational Mathematics, 
Princeton University, Princeton, NJ 08544, USA}
\author{Jiequn Han}
\affiliation{Program in Applied and Computational Mathematics, 
Princeton University, Princeton, NJ 08544, USA}
\author{Weinan E}
\affiliation{Department of Mathematics and Program 
in Applied and Computational Mathematics, 
Princeton University, Princeton, NJ 08544, USA}
\affiliation{Center for Data Science, 
Beijing International Center for Mathematical Research,  Peking University,
Beijing Institute of Big Data Research, 
Beijing, 100871, P.R.~China}

\begin{abstract}
Recent developments in many-body potential energy representation via deep learning
have brought new hopes to addressing the accuracy-versus-efficiency dilemma in molecular simulations.
Here we describe DeePMD-kit, 
a package written in Python/C++ that has been designed to minimize the effort required to 
build  deep learning based representation of potential energy and force field and to perform molecular dynamics. 
Potential applications of DeePMD-kit span from finite molecules to extended systems
and from metallic systems to chemically bonded systems.
 DeePMD-kit is interfaced with TensorFlow, one of the most popular deep learning frameworks, making
 the training process  highly automatic and efficient.
 On the other end, DeePMD-kit is interfaced with  high-performance classical molecular dynamics 
 and quantum (path-integral) molecular dynamics packages, i.e.,~LAMMPS and the i-PI, respectively.
Thus, upon training, the potential energy and force field models can be used to perform efficient molecular simulations for different purposes.
As an example of the many potential applications of the package, 
we use DeePMD-kit to learn the interatomic potential energy and forces of a water model using  data obtained from
 density functional theory. We demonstrate that the resulted  
molecular dynamics model  reproduces accurately the structural information contained in the original model.

\end{abstract}

\maketitle

\setlength{\parskip}{.5em}

\section{Introduction}
The dilemma of accuracy versus efficiency in modeling the potential energy surface (PES) and interatomic forces 
has confronted the  molecular simulation communities
for a long time.
On one hand,
{\it ab initio} molecular dynamics (AIMD) has the accuracy of the 
density functional theory (DFT)~\cite{kohn1965self,car1985unified,marx2009ab},
but  the computational cost of DFT in evaluating the PES and forces  
restricts its typical applications to system size of hundreds to thousands of atoms and time scale of $\sim$ 100 $ps$. 
One the other hand, a great deal of effort has been made in developing empirical force fields (FFs)
\cite{vanommeslaeghe2010charmm,jorgensen1996development,wang2004development},
which allows for much larger and longer simulations.
However, the accuracy and transferability of FFs is often in question.
Moreover, fitting the parameters of an FF  is usually a tedious and ad hoc process.

\newparagraph
In the last few years,  machine learning methods have been suggested as a tool to model PES of molecular systems with DFT data,
and has achieved some remarkable success
\cite{thompson2015SNAP,huan2017AGNI,behler2007generalized, morawietz2016van,bartok2010gaussian,rupp2012fast,
schutt2017quantum,chmiela2017machine,smith2017ani,yao2017tensormol}.
Some 
examples (not a comprehensive list)
include the Behler-Parrinello neural network (BPNN)~\cite{behler2007generalized},
the Gaussian approximation potentials (GAP)~\cite{bartok2010gaussian},
the Gradient-domain machine learning (GDML)~\cite{chmiela2017machine},
and the Deep potential for molecular dynamics (DeePMD)~\cite{han2017deep,zhang2017deep}.
In particular, it has been demonstrated for a wide variety of systems
that the ``deep potential'' and DeePMD allow us to perform
molecular dynamics simulation with accuracy comparable to that of DFT (or other fitted data) and  the efficiency competitive with
empirical potential-based molecular dynamics \cite{han2017deep,zhang2017deep}.

\newparagraph
Machine learning, particularly deep learning has been shown to be a powerful tool in a variety of fields~\cite{lecun2015deep, goodfellow2016deep}
and even has outperformed human experts in some applications like the AlphaGo in the board game Go~\cite{silver2016mastering}.
A number of open source deep learning platforms, e.g. TensorFlow~\cite{abadi2016tensorflow}, Caffe~\cite{jia2014caffe}, Torch~\cite{collobert2011torch7}, and MXNet~\cite{chen2015mxnet} are available.  These open source platforms have significantly lowered the technical barrier for the application of deep learning.
Considering the potential impact that deep learning-based methods will have on molecular simulation, it 
 is of considerable interest to develop open source platforms that serve as the  interface  between deep neural network models and molecular simulation tools such as LAMMPS~\cite{plimpton1995fast}, Gromacs~\cite{hess2008gromacs} and NAMD~\cite{phillips2005scalable},
 and path-integral MD packages like i-PI~\cite{ceriotti2014ipi}.

\newparagraph

\newparagraph
The contribution of this work is to provide an implementation of the DeePMD method,
namely DeePMD-kit,
which interfaces with TensorFlow for fast training, testing, and evaluation of the PES and forces,
and with LAMMPS and i-PI for classical and path-integral molecular dynamics simulations, respectively. 
In DeePMD-kit,
we implement the atomic environment descriptors
and chain rules for force/virial computations in C++
and provide an interface to incorporate them as new operators in  standard TensorFlow.
This allows the model training and MD simulations to benefit from TensorFlow's highly optimized tensor operations.
The support of DeePMD for LAMMPS is implemented as a new ``pair style'',
the standard command in LAMMPS.
Therefore, only a slight modification
in the standard LAMMPS input script is required for energy, force, and virial evaluation through DeePMD-kit.
The support for i-PI is implemented as a new force client communicating through sockets with the standard i-PI server, which handles the bead integrations.
Given these features provided by DeePMD-kit,
training deep neural network model for potential energy and running MD simulations with the model 
is made much easier than implementing everything from scratch.


\newparagraph
The manuscript is organized as follows.
In section~\ref{sec:theory}, the theoretical framework of the DeePMD method is provided.
We show in detail how the system energy is constructed
and how to take derivatives with respect to the atomic position and box tensor to compute the force and virial.
In section~\ref{sec:software}, we provide a brief introduction on how to use DeePMD-kit to train a model
and run MD simulations with the model.
In section~\ref{sec:example}, we demonstrate the performance of DeePMD-kit by
training a DeePMD model from AIMD data.
Results from the MD simulation using the trained DeePMD model are compared to the original AIMD data to validate the modeling.
The paper  concludes with a discussion about the future work planed for DeePMD-kit.

\section{Theory}
\label{sec:theory}

We consider a system consisting of $N$ atoms and denote the coordinates of the atoms by $\{\vect R_1, \dots, \vect R_N\}$.
The potential energy $E$ of the system is a function with $3N$ variables,
i.e.,~$E = E(\vect R_1, \dots, \vect R_N)$, with each $\vect R_i \in \mathbb R^3$.
In the DeepMD method, $E$ is decomposed into a sum of atomic energy contributions,
\begin{align}
  E = \sum_i E_i,
\end{align}
with $i$ being the indexes of the atoms.
Each atomic energy is fully determined by the position of the $i$-th atom and its near neighbors,
\begin{align}\label{eqn:atom-ener}
  E_i = E_{s(i)} (\vect R_i, \{\vect R_j \,\vert\, j\in \neighbor_{R_c}(i)\}),
\end{align}
where $\neighbor_{R_c}(i) $ denotes the index set of the neighbors of atom $i$ within the cut-off radius $R_c$,
i.e.~$R_{ij} = \vert \vect R_{ij}\vert  = \vert \vect R_i  - \vect R_j \vert \leq R_c$.
$s(i)$ is the chemical specie of atom $i$.
The most straightforward idea to model the atomic energy $ E_{s(i)}$ through DNN is to train a neural network with the input simply 
being the positions of the $i$th atom $\vect R_i$ and its neighbors $\{\vect R_j \,\vert\, j\in \neighbor_{R_c}(i)\}$.
This approach is less than optimal 
as it does not guarantee the translational, rotational, and permutational symmetries lying in the PES. 
Thus, a proper preprocessing of the atomic positions, which maps the positions 
to ``descriptors'' of atomic chemical 
environment~\cite{bartok2013representing} is needed.

\newparagraph
In the DeePMD method, to construct the descriptor for atom $i$, the positions of its neighbors are firstly shifted by the position of atom $i$, viz.~$\vect R_{ij} = \vect R_i - \vect R_j$.
The coordinate of the relative position $\vect R_{ij}$  under lab frame $\{\vect e^0_x, \vect e^0_y, \vect e^0_z\}$ is denoted by 
$(x^0_{ij}, y^0_{ij}, z^0_{ij})$, i.e.,
\begin{align}\label{eqn:global-xyz}
  \vect R_{ij} = x^0_{ij} \vect e^0_x  + y^0_{ij} \vect e^0_y + z^0_{ij} \vect e^0_z.
\end{align}
Both $\vect R_{ij}$  and the coordinate $(x^0_{ij}, y^0_{ij}, z^0_{ij})$ preserve the translational symmetry.
The rotational symmetry is preserved by constructing a local frame and recording the local coordinate for each atom.
First, two atoms, indexed  $a(i)$ and $b(i)$,  are picked from the neighbors $N_{R_c}(i)$ by certain user-specified rules.
The local frame $\{\vect e_{i1}, \vect e_{i2}, \vect e_{i3}\}$ of atom $i$ is then constructed by 
\begin{align} \label{eqn:e1}
  \vect e_{i1} &= \vect e (\vect R_{ia(i)}), \\
  \vect e_{i2} &= \vect e \big(\vect R_{ib(i)} - (\vect R_{ib(i)} \cdot \vect e_{i1}) \vect e_{i1}\big), \\\label{eqn:e3}
  \vect e_{i3} &= \vect e_{i1} \times \vect e_{i2},
\end{align}
where $\vect e(\vect R)$ denotes the normalized vector of $\vect R$, i.e., $\vect e(\vect R) = \vect R / \vert \vect R\vert$.
Then the local coordinate $(x_{ij}, y_{ij}, z_{ij})$ (under the local frame)
is transformed from the global coordinate $(x^0_{ij}, y^0_{ij}, z^0_{ij})$ through
\begin{align}\label{eqn:rot}
  (x_{ij}, y_{ij}, z_{ij}) = (x^0_{ij}, y^0_{ij}, z^0_{ij}) \cdot \mr ( \vect R_{ia(i)}, \vect R_{ib(i)} ),
\end{align}
where
\begin{align}
  \mr ( \vect R_{ia(i)}, \vect R_{ib(i)} ) = [\vect e_{i1}, \vect e_{i2}, \vect e_{i3}]  
\end{align}
is the rotation matrix with the columns being the local frame vectors.
The descriptive information of atom $i$ given by neighbor $j$ is constructed by using either full information (both radial and angular) or radial-only information:
\begin{align}\label{eqn:dija}
  \{D_{ij}^\alpha\}
  = \left\{
  \begin{aligned}
   & \Big\{\frac 1 { R_{ij}}, \frac{x_{ij}}{ R_{ij}} , \frac{y_{ij}}{ R_{ij}} , \frac{z_{ij}}{ R_{ij}}\Big\},   & &\text{full information;} \\
   & \Big\{\frac 1 { R_{ij}}\Big\},   & &\text{radial-only information.} \\
  \end{aligned}
        \right.
\end{align}
When $\alpha=0,1,2,3$, full (radial plus angular) information is provided.
When $\alpha=0$, only radial information is used.
It is noted that the order of the neighbor indexes $j$'s in $\{D_{ij}^\alpha\}$ is fixed by
sorting them firstly according to their chemical species
and then, within each chemical species, according to their inversed distances to atom $i$, i.e., $1/R_{ij}$.
The permutational symmetry is naturally preserved in this way.
Following the aforementioned procedures, we have constructed the mapping from atomic positions to descriptors, 
which is denoted by
\begin{align}\label{eqn:step1}
  \vect D_i =  \vect D_i (\vect R_i, \{\vect R_j \,\vert\, j\in \neighbor_{R_c}(i)\}). 
\end{align}
The components
$  D_{ij}^\alpha = D_{ij}^\alpha (\vect R_{ij}, \vect R_{ia(i)}, \vect R_{ib(i)})$
are given by Eqns.~\eqref{eqn:global-xyz}--\eqref{eqn:dija}.
The descriptors
$\vect D_i$ preserve the translational, rotational, 
and permutational symmetries
and are passed to a DNN to evaluate the atomic energy.
We refer to the Supplementary Materials of Ref.~\cite{zhang2017deep} for further details in selection of
axis atoms and standardization of input data.

\newparagraph
The DNN that maps the descriptors $\vect D_i$ to atomic energy is denoted by
\begin{align}\label{eqn:step2}
  E_{s(i)}
  = \mn_{s(i)} (\vect D_i).
\end{align}
It is a feedforward network
in which data flows from the input layer as $\vect D_i$, through multiple fully connected hidden layers, to the output layer as the atomic energy $E_{s(i)}$.
Mathematically, DNN with $N_h$ hidden layers is a mapping
\begin{align}
  \mn_{s(i)} (\vect D_i)
  = \ml^\olayer_{s(i)} \circ \ml_{s(i)}^{N_h} \circ \ml_{s(i)}^{N_h-1} \circ \cdots \circ \ml_{s(i)}^1 (\vect D_i),
\end{align}
where the symbol ``$\circ$'' denotes function composition.
Here $ \ml_{s(i)}^{p} $ is the mapping from layer $p-1$ to $p$, which is a composition of a linear transformation and a non-linear transformation, 
the so-called activation function:
\begin{align}\label{eqn:layer}
  \vect d_i^p = \ml_{s(i)}^p (\vect d_i^{p-1}) = \varphi \big( \vect W_{s(i)}^p  \vect d_i^{p-1}  + \vect b_{s(i)}^p   \big),
\end{align}
where $\vect d^p_i \in \mathbb R^{M_p}$ denotes the value of neurons in layer $p$ and $M_p$ the number of neurons.
The weight matrix $\vect W_{s(i)}^p \in \mathbb R^{M_p\times M_{p-1}}$ and bias vector $\vect b_{s(i)}^p \in \mathbb R^{M_p}$
are free parameters of the linear transformation that are to be optimized.
The non-linear activation function $\varphi$ is in general a component-wise function,
and here it is taken to be the hyperbolic tangent, i.e.,
\begin{align}
  \varphi (d_1, d_2, \dots, d_M) = (\tanh(d_1), \tanh(d_2), \dots, \tanh(d_M)).
\end{align}
The output mapping $\ml^\olayer_{s(i)}$ is a linear transformation
\begin{align}
  E_{s(i)} = \ml^\olayer_{s(i)} (\vect d^{N_h}_i) =  \vect W_{s(i)}^\olayer  \vect d^{N_h}  + b_{s(i)}^\olayer,
\end{align}
where weight vector $ \vect W_{s(i)}^\olayer \in \mathbb R^{1\times M_{N_h}}$  and bias $b_{s(i)}^\olayer\in \mathbb R$ are free parameters to be optimized as well.

\newparagraph
The force on the $i$-th atom is computed by taking the negative gradient of the system energy with respect to its position,
which is given by
\begin{equation}\label{eqn:force}
\begin{split}
  \vect F_i
  =&\,
    -\sum_{j\in N(i), \alpha}\frac{\partial \mn_{s(i)}}{\partial D_{ij}^\alpha} \frac{\partial D_{ij}^\alpha}{\partial \vect R_i} -\sum_{j\neq i}\sum_{k\in N(j),\alpha}\delta_{i,a(j)}\frac{\partial \mn_{s(j)}}{\partial D_{jk}^\alpha} \frac{\partial D_{jk}^\alpha}{\partial \vect R_i}\\ 
   &\,-\sum_{j\neq i}\sum_{k\in N(j),\alpha}\delta_{i,b(j)}\frac{\partial \mn_{s(j)}}{\partial D_{jk}^\alpha} \frac{\partial D_{jk}^\alpha}{\partial \vect R_i} 
   -\sum_{j\neq i}\sum_{k\in \tilde N(j),\alpha}\delta_{i,k}\frac{\partial \mn_{s(j)}}{\partial D_{jk}^\alpha} \frac{\partial D_{jk}^\alpha}{\partial \vect R_i},
\end{split}
\end{equation}
where $\tilde N(j) = N(j) - \{a(j), b(j)\}$. 
The virial of the system is given by
\begin{equation}\label{eqn:virial}
\begin{split}
  \Xi
  =&\,
      \frac12 \sum_{i\neq j} \vect R_{ij} \sum_{\alpha}\frac{ \partial \mn_{s(i)}}{\partial D_{ij}^\alpha} \frac{\partial D_{ij}^\alpha}{\partial \vect R_{ij}}    
     + \frac12 \sum_{i\neq j} \delta_{j,a(i)} \vect R_{ij} \sum_{q,\alpha} \frac{ \partial \mn_{s(i)}}{\partial D_{iq}^\alpha} \frac{\partial D_{iq}^\alpha}{\partial \vect R_{ij}} \\
&\,   + \frac12 \sum_{i\neq j} \delta_{j,b(i)} \vect R_{ij} \sum_{q,\alpha} \frac{ \partial \mn_{s(i)}}{\partial D_{iq}^\alpha} \frac{\partial D_{iq}^\alpha}{\partial \vect R_{ij}}.
\end{split}
\end{equation}
The derivation of the force and virial formula Eqs. \eqref{eqn:force}--\eqref{eqn:virial} is given in~\ref{app:tmp1}.

\newparagraph
The unknown parameters $\{\vect W_s^p, \vect b_s^p\}$ in the linear transformations of the DNN are determined 
by a {training} process that minimizes the \emph{loss function} $L$, i.e.,
\begin{align}\label{eqn:opti-prob}
  \min_{\{\vect W_s^p,\vect b_s^p\}} L(p_\epsilon, p_f, p_\xi).
\end{align}
The $L$ is defined as a sum of different mean square errors of the DNN predictions
\begin{align}\label{eqn:loss}
  L(p_\epsilon, p_f, p_\xi) = \frac{p_\epsilon}{N} \Delta E^2 + \frac{p_f}{3N} \sum_i\vert\Delta\vect F_i\vert^2 + \frac{p_\xi}{9N} \Vert\Delta \Xi\Vert^2,
\end{align}
where $\Delta E$, $\Delta \vect F_i$ and $\Delta\Xi$ denotes root mean square (RMS) error in energy, force, and virial, respectively.
The prefactors $p_\epsilon$, $p_f$, and $p_\xi$ are free to change even during the optimization process.
In this work, the prefactors are given by
\begin{align}
  p(t) = p^\limit\Big [ 1 - \frac{r_l(t)}{r_l^0} \Big]  + p^\start\Big [\frac{r_l(t)}{r_l^0} \Big],
\end{align}
where $r_l(t)$ and $r_l^0$ are the learning rate at training step $t$ and the learning rate at the beginning, respectively. 
The prefactor varies from $p^\start$ at the beginning and goes to $p^\limit$ as the learning ends.
We adopt an exponentially decaying learning rate 
\begin{align}\label{eqn:lr}
  r_l(t) = r_l^0 \times d_r ^ {\ t/d_s},
\end{align}
where $d_r$ and $d_s$ are the decay rate and decay steps, respectively. The decay rate $d_r$ is required to be less than 1.

\section{Software}
\label{sec:software}

\begin{figure}[]
  \centering
  \includegraphics[width=0.49\textwidth]{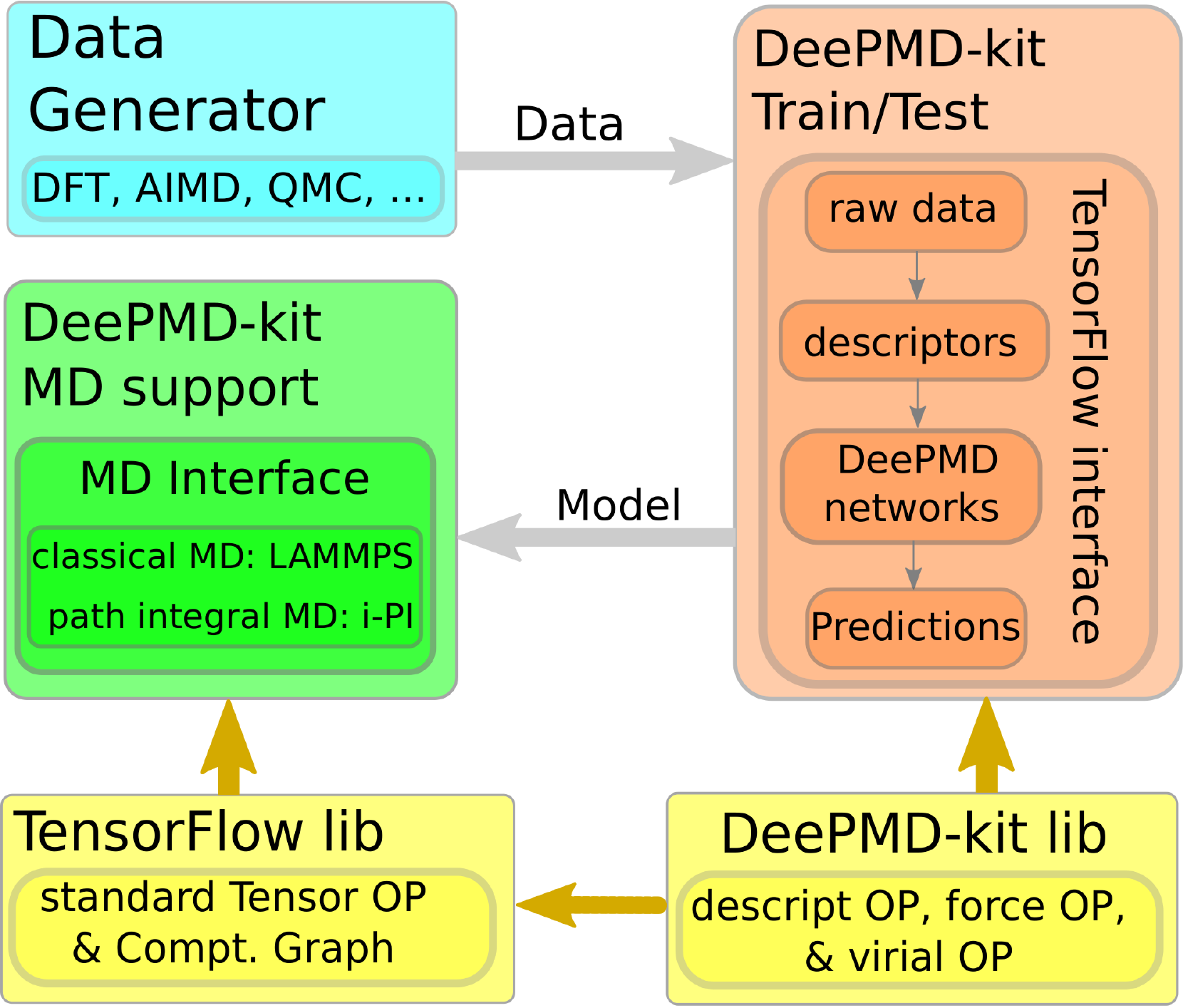}
  \caption{Schematic plot of the DeePMD-kit architecture and the workflow. 
  The gray arrows present the workflow.
  The data, including energy, force, virial, box, and type, 
  are passed from the Data Generator to the DeePMD-kit Train/Test module to perform training. 
  After training, the DeePMD model is passed to the DeePMD-kit MD support module to perform MD.
  The TensorFlow and DeePMD-kit libraries are used for supporting different calculations.
  See text for detailed descriptions.
  }
  \label{fig:flowchart}
\end{figure}

The DeePMD-kit is composed of three parts:
(1) a library that implements the computation of descriptors, forces, and virial in C++, including interfaces to TensorFlow and third-party MD packages;
(2) training and testing programs built on TensorFlow's Python API;
(3) supports for LAMMPS and i-PI.
This section illustrates the usage of DeePMD-kit along a typical workflow:
preparing data, training the model, testing the model, and running classical/path-integral MD simulations with the model.
A schematic plot of the DeePMD-kit architecture and the workflow is shown in Fig.~\ref{fig:flowchart}.

\subsection{Data preparation}


The data for training/testing a DeePMD model is composed of a list of \verb|systems|.
Each \verb|system| contains a number of \verb|frames|.
Some of the \verb|frames| are used  as training data, while the others are used as testing data. 
Each \verb|frame| records the shape of simulation region (box tensor) and the positions of all atoms in the system.
The order of the \verb|frames| in a \verb|system| is not relevant, 
but the number of atoms and the atom types should be the same for all \verb|frames| in the same \verb|system|.
Each \verb|frame| is labeled with the energy, the forces, and the virial.
Any one or two of the labels can be absent.
When a label is absent, its corresponding prefactor in the loss function Eq.~\eqref{eqn:loss} is set to zero.
The labels can be computed by any molecular simulation package that takes in the atomic positions and the box tensor
and returns the energy, the forces, and/or the virial.
The DeePMD-kit defines a data protocol called RAW format. 
The labels computed by different packages should be converted to RAW format to serve as training/testing data.
The box tensor, atomic coordinates (under lab frame), and the labels are stored in separate text files,
with names \texttt{box.raw}, \texttt{coord.raw}, \texttt{energy.raw}, \texttt{force.raw}, and \texttt{virial.raw}, respectively.
Each line of a RAW file corresponds to one \verb|frame| of the data, with the properties of each atom presented in succession.
The order of the \verb|frames| appearing in a RAW files and the order of atoms in each \verb|frame| should be consistent across all the RAW files.
Taking \texttt{coord.raw} as an example,
the first three numbers in the first line are the coordinate of the first atom in the first \verb|frame|,
the next three numbers are the coordinate of the second atom, and so forth.
The first three numbers in the second line are the coordinate of the first atom in the second \verb|frame|.
The units of length, energy, and force in the RAW files are \AA, eV, and eV/\AA, respectively.
The data is organized in this way because the \verb|frames| can be combined or split
in a convenient way using standard text processing tools such as \texttt{cat}, \texttt{sed}, and \texttt{awk} provided by Unix-like operating systems, 
and the files can also be manipulated and analyzed as array text data by the NumPy module of Python.
The atom types are recorded in the file \texttt{type.raw}, which has only one line with atom types as integers presented in succession.
Again, it is addressed that the atom types should be consistent in all \verb|frames| of the same \verb|system|.

\newparagraph
The data is composed of several \verb|systems|.
The RAW files of the different \verb|systems| should be placed in different folders, and
the number of atoms and the atom types are NOT required to be the same for different \verb|systems|.
Frequent loading of the RAW text files from hard disk may become the bottleneck of efficiency. 
Therefore, the RAW files except the  \texttt{type.raw} are firstly converted to NumPy binary files
and then used by the training and testing programs in DeePMD-kit.
DeePMD-kit provides a Python script for this conversion.

\subsection{Model training}
\label{sec:training}

The computation of atomic energy $E_{s(i)}$ (see Eq.~\eqref{eqn:atom-ener}) consists of two successive mappings:
first, from the positions of the atom $i$ and its neighbors to its descriptors, 
i.e., Eq.~\eqref{eqn:step1}; 
second, from the descriptors to the atomic energy through DNN, i.e.,~Eq.~\eqref{eqn:step2}.
The DNN part is implemented by standard tensor operations provided by the TensorFlow deep learning framework.
However, the descriptor part is not a standard operation in TensorFlow,
thus it is implemented with C++ and is interfaced to TensorFlow as a new ``operator''.
The force and virial computation requires derivatives of system energy 
with respect to atomic position and box tensor, respectively.
This is done with the chain rule in Eq.~\eqref{eqn:force} and Eq.~\eqref{eqn:virial}, respectively.
The gradient of the DNN, i.e.,~$\partial E_{s(i)} / \partial D_{jk}^\alpha$, is implemented by the \verb|tf.gradients| operator provided by TensorFlow.
The derivatives $\partial D_{jk}^\alpha/ \partial \vect R_l$ and the chain rules defined in Eq.~\eqref{eqn:force} and Eq.~\eqref{eqn:virial}
are implemented in C++ and then interfaced with TensorFlow.
By using the TensorFlow with the user implemented operators, we are now able to compute 
the system energy, the atomic forces, and the virial,
thus we are able to evaluate the loss function (forward propagation).
The derivatives of the loss function with respect to the parameters 
$\{\vect W_s^p,\vect b_s^p\}$ (backward propagation) are automatically computed by TensorFlow.

\newparagraph
The optimization problem~\eqref{eqn:opti-prob} is currently solved
by the TensorFlow's implementation of the Adam stochastic gradient descent method~\cite{kingma2014adam}.
At each step of optimization (equivalent to training step),
the value and gradients of the loss function is computed against only a subset of the training data, which is called a \emph{batch}.
The number of \verb|frames| in a batch is called the \emph{batch size}.
Taking the RMS energy error $\Delta E$ for instance, it is evaluated by
\begin{align}
  \Delta E^2 = \frac{1}{S_b} \sum_{k=1}^{S_b} \vert E^k - E (\vect R^k_1, \dots,\vect R^k_N)\vert^2
\end{align}
where $\{\vect R^k\}$, $E^k$, and $S_b$ denote the atomic positions, system energy of the $k$th frame in the batch, and the batch size, respectively. 
The  errors $\vert\Delta \vect F_i\vert^2$ and $\Vert\Delta\Xi\Vert^2$ are evaluated analogously.
It is noted that the evaluation of the loss function for different frames in the batch is embarrassingly parallel.
Therefore, ideally, the batch size $S_b$ should be divisible by the number of CPU cores in the computation.

\newparagraph
We denote the \verb|systems| in the training data by $\{\Omega_1, \dots, \Omega_{S_s}\}$ with $S_s$ being the total number of systems 
and denote the number of frames in $\Omega_i$ by $\vert \Omega_i\vert$. 
The \verb|systems| $\{\Omega_1, \dots, \Omega_{S_s}\}$ are used in the training in a cyclic way.
First, the model is trained for $\vert \Omega_1\vert /S_b$ steps by using $\vert \Omega_1\vert /S_b$ batches randomly taken from $\Omega_1$ without replacement.
Next, the model is trained for $\vert \Omega_2\vert /S_b$ steps by using $\vert \Omega_2\vert /S_b$ batches randomly taken from $\Omega_2$ without replacement.
In such a way, the \verb|systems| in the set $\{\Omega_1, \dots, \Omega_{S_s}\}$ are used in training successively.

\newparagraph
The training program in the DeePMD-kit is called \verb|dp_train|.
It reads a parameter file in JSON format that specifies the training process.
Some important settings in the parameter file are
\begin{verbatim}
{
    "n_neuron":    [240, 120, 60, 30, 10],
    "systems":     ["/path/to/water", "/path/to/ice"],
    "stop_batch":  1000000,
    "batch_size":  4,
    "start_lr":    0.001,
    "decay_steps": 5000,
    "decay_rate":  0.95,
}
\end{verbatim}
In this file, the item \verb|n_neuron| sets the number of hidden layers to 5, and the number of neurons in each layer are set to 
$(M_1, M_2, M_3, M_4, M_5) = (240, 120, 60, 30, 10)$, from the innermost to the outermost layer.
The training has two \verb|systems|, with $\Omega_1$ stored in the folder \verb|/path/to/water| and $\Omega_2$ in the folder \verb|/path/to/ice|.
The batch size is set to 4.
In total the model is optimized for $10^6$ steps (set by \verb|stop_batch|), i.e., $10^6$ batches are used in the training.
The starting learning rate, decay steps, and decay rate (see Eq.~\eqref{eqn:lr}) are set to 0.001, 5000, and 0.95, respectively.

\newparagraph
The parameters of the DeePMD model is saved to TensorFlow checkpoints during the training process,
thus one can break the training at any time and restart it from any of the checkpoints. 
Once the training finishes, the model parameters and the network topology are \emph{frozen} from the checkpoint file by the tool \verb|dp_frz|.
The frozen model can be used in model testing and MD simulations. 

\subsection{Model testing}
\begin{figure}
  \centering
  \includegraphics[width=0.5\textwidth]{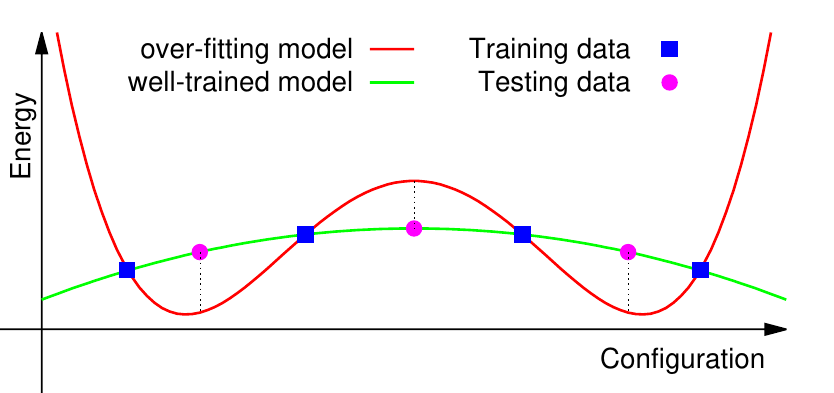}
  \caption{Schematic illustration of over-fitting.
    The blue squares denote the training data, while the pink filled circles denote the testing data.
    Only the training data is used in training models.
    Both the over-fitting model and the well-trained model have small training error,
    however, the over-fitting model presents a significantly larger testing error.
  }
  \label{fig:of}
\end{figure}

DeePMD-kit provides two modes of model testing.
(1) During the training,
the RMS energy, force and virial errors and the loss function are evaluated by both the training batch data and the testing data
and displayed on the fly.
Sometimes, for the sake of efficiency, only a subset of the testing data is used to test the model on the fly.
(2) After the model is frozen, it can be tested by the tool \verb|dp_test|.
Ideally the training error and the testing error should be roughly the same.
A signal of overfitting is indicated by a much lower training error compared to the testing error, 
see Fig.~\ref{fig:of} for an illustration.
In this case, it is suggested to either reduce the number of layers 
and/or the number of neurons of each layer, or increase the size of the training data.


\subsection{Molecular dynamics}

Once the model parameters are frozen, MD simulations can be carried out.
We provide an interface that inputs the atom types and positions and returns energy, forces, and virial computed by the DeePMD model.
Therefore, in principle, it can be called in any MD package during MD simulations. 
In the current release of DeePMD-kit, we provide supports for the LAMMPS and i-PI packages.

\newparagraph
The evaluation of interactions is implemented by using the TensorFlow's C++ API. 
First, the model parameters are loaded, then the network operations defined in the frozen model are executed
in exactly the same way as the 
evaluation in the model training stage, see Sec.~\ref{sec:training}.
The DeePMD-kit's implementation of descriptors, derivatives of descriptors, chain rules for force and virial computations
are called as non-standard operators by the TensorFlow.

\newparagraph
\paragraph* {LAMMPS support}
The LAMMPS support for DeePMD is shipped as a third-party package with the DeePMD-kit source code.
The installation of package is similar to other third-party packages for LAMMPS and is explained in detail in the DeePMD-kit manual.
In the current release, only serial MD simulations with DeePMD model are supported. 
To enable the DeePMD model, only two lines are added in the LAMMPS input file. 
\begin{verbatim}
pair_style     deepmd graph.pb
pair_coeff     
\end{verbatim}
The command \verb|deepmd| in \verb|pair_style| means to use the DeePMD model to compute the atomic interactions in the MD simulations.
The parameter \verb|graph.pb| is the file containing the frozen model.
The \verb|pair_coeff| should be left blank.

\newparagraph
\paragraph* {i-PI support} The i-PI is implemented based on a client-server model.
The i-PI works as a server that integrates the trajectories of the nucleus.
The DeePMD-kit provides a client called \verb|dp_ipi| that gets coordinates of atoms from the i-PI server
and returns the energy, forces, and virial computed by the DeePMD model to the i-PI server.
The communication between the server and client is implemented through either the UNIX domain sockets or the Internet sockets.
It is noted that multiple instances of the client is allowed, thus the computation of the interactions in multiple path-integral replicas is embarrassingly parallelized.
The parameters of running the client are provided by a JSON file. An example for a water system is
\begin{verbatim}
{
    "verbose":          false,
    "use_unix":         true,
    "port":             31415,
    "host":             "localhost",
    "graph_file":       "graph.pb",
    "coord_file":       "conf.xyz",
    "atom_type" :       {"OW": 0, "HW1": 1, "HW2": 1}
}
\end{verbatim}
In this example, the client communicates with the server through the UNIX domain sockets at port \verb|31415|.
The forces are computed according to the frozen model stored in \verb|graph.pb|.
The \verb|conf.xyz| file provides the atomic names and coordinates of the system.
The \verb|dp_ipi| ignores the coordinates in \verb|conf.xyz| and translates the atom names to types according to the rule provided by \verb|atom_type|.

\section{Example}
\label{sec:example}
The performance of the DeePMD-kit package is demonstrated by a bulk liquid water system of 64 molecules subject to periodic boundary conditions\footnote{For this example, the raw data and the JSON parameter files for training and MD simulation are provided in the online package.
More details on how to use them are explained in the manual.}.
The dataset is generated by a 20 ps, 330 K NVT AIMD simulation with PBE0+TS exchange-correlation functional.
The frames are recorded from the trajectory in each time step, i.e., 0.0005 ps. 
Thus in total we have 40000 frames.
The order of the frames is randomly shuffled. 
38000 of them are used as training data,
while the remaining 2000 are used as testing data. 

The cut-off radius of neighbor atoms is $6.0$~\AA.
The network input contains both radial and angular information of 16 closest neighboring oxygen atoms and 32 closest neighboring hydrogen atoms,
while contains only the radial information of the rest of neighbors.
The DNN contains 5 hidden layers.
The size of each layer is $(M_1, M_2, M_3, M_4, M_5) = ( 240, 120, 60, 30, 10)$, from the innermost to the outermost layer.
The model is trained by the Adam stochastic gradient descent method, with the learning rate decreasing exponentially.
The decay rate and decay step are set to 0.95 and 5000, respectively.
The prefactors in the loss function are taken as
$p^\start_e = 0.02$,
$p^\limit_e = 8$, 
$p^\start_f = 1000$, 
and $p^\limit_f = 1$. 
No virial is available in the data, 
so the virial prefactors are set to 0, i.e., $p^\start_v = p^\limit_v = 0$.

\begin{figure}
  \centering
  \includegraphics[width=0.45\textwidth]{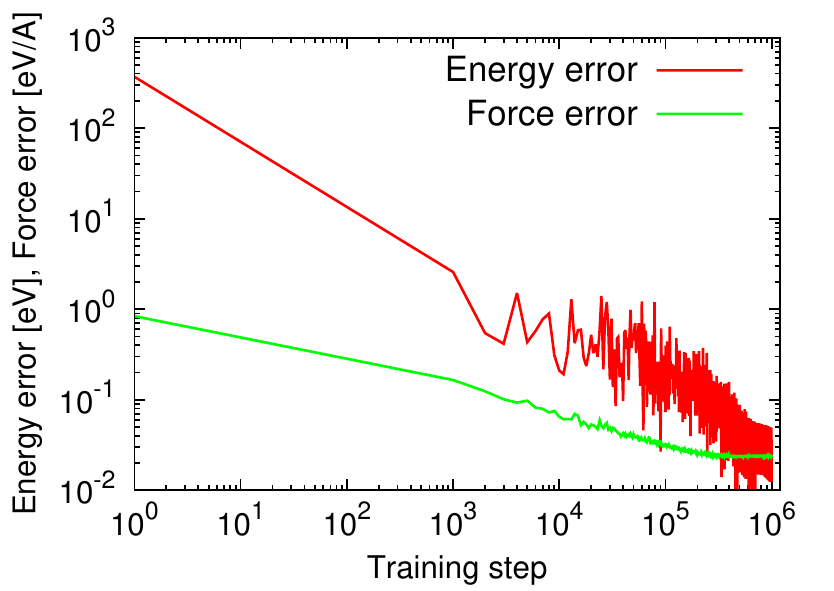}
  \caption{The learning curves of the liquid water system.
    The root mean square energy and force testing errors are presented against the training step.
    The energy error is given in the unit of eV, while the force error is given in unit of eV/\AA. The axes of the plot are logscaled.}
  \label{fig:lc}
\end{figure}

\newparagraph
The model is trained on a desktop machine with an Intel Core i7-3770 CPU and 32GB memory using 4 OpenMP threads.
The total wall time of the training is 16 hours.
The learning curves of the RMS energy and force errors as functions of training step are plotted in Fig.~\ref{fig:lc}.
The errors are tested on the fly by 100 frames randomly picked from the testing set.
At the beginning, the model parameters $\{\vect W_s^p, \vect b_s^p\}$ are randomly initialized,
and the RMS energy and force errors are $3.7\times 10^2$~eV and $8.4\times 10^{-1}$~eV/\AA, respectively. 
At the end of training, the RMS energy and force errors over the whole testing set are $2.8\times 10^{-2}$~eV and $2.4\times 10^{-2}$~eV/\AA, respectively.
The standard deviation of the energy and the forces in the data 
are $6.5\times 10^{-1}$~eV and $8.1\times 10^{-1}$~eV/\AA, respectively.
Therefore, the relative errors of energy and force with respect to
the data standard deviation are 4.3\% and 2.9\%, respectively.


\begin{figure}
  \centering
  \includegraphics[width=0.45\textwidth]{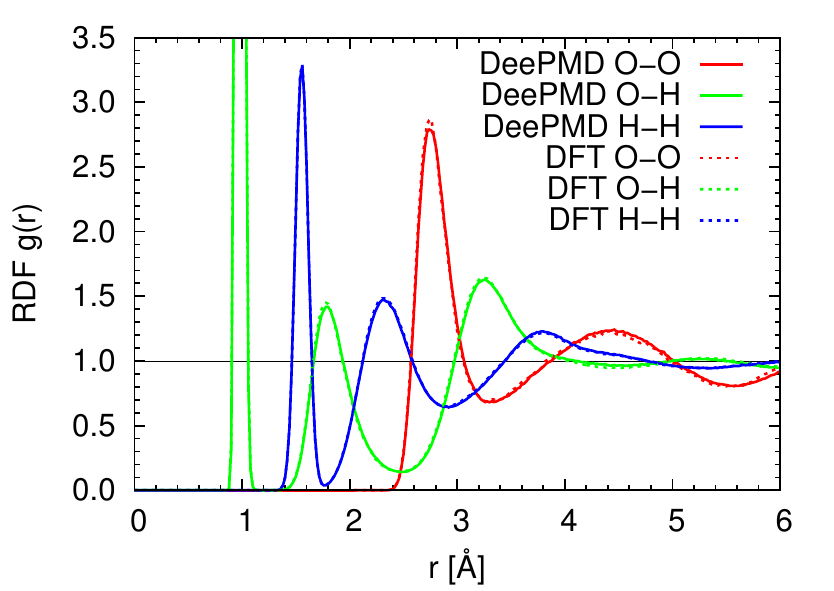}
  \caption{The radial distribution functions of the DeePMD compared with the PBE0+TS DFT water model.}
  \label{fig:rdf}
\end{figure}

\begin{figure}
  \centering
  \includegraphics[width=0.45\textwidth]{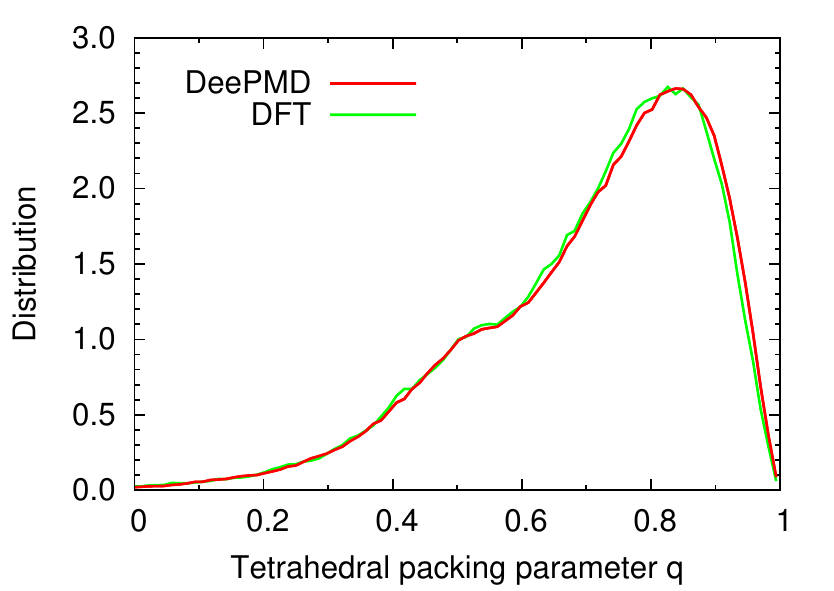}
  \caption{The distribution of the tetrahedral packing parameter of the DeePMD compared with the PBE0+TS DFT water model.}
  \label{fig:tp}
\end{figure} 

\newparagraph
The trained DeePMD model is frozen and passed to LAMMPS to run NVT MD simulation of 64 water molecules.
The simulation cell is of size $12.4447\,\textrm{\AA}\times 12.4447\,\textrm{\AA} \times 12.4447\,\textrm{\AA}$
under periodic boundary conditions.
The simulation lasts for 200~ps. 
Snapshots in the first 50~ps are discarded,
while the rest snapshots in the trajectory is saved in every other 0.01~ps for structural analysis.
The oxygen-oxygen, oxygen-hydrogen, and hydrogen-hydrogen radial distribution functions are presented in Fig.~\ref{fig:rdf}.
The distribution of the tetrahedral packing parameter~\cite{errington2001relationship} is presented in Fig.~\ref{fig:tp}.
These results show that the DeePMD model is in satisfactory agreement with the DFT model
in generating structure properties.

\section{Conclusion and future work}
\label{sec:conclusion}

We introduced the software DeePMD-kit, which implements DeePMD, a deep neural network representation for atomic interactions, based on the deep learning framework TensorFlow.
The descriptors and chain rules for force/virial computation of DeePMD is implemented in C++ and interfaced to TensorFlow as new operators for model training and PES evaluation.
Therefore, the training, testing, and MD simulations benefit from 
TensorFlow's state-of-the-art training algorithms and highly optimized tensor operations.
Supports for third-party MD packages, LAMMPS and i-PI, are provided such that these softwares can do classical/path-integral MD simulations with the atomic interactions modeled by DeePMD.

\newparagraph
In addition, we also provided the analytical details needed to implement the DeePMD method,
including the definition of the chemical environment descriptors, the deep neural network architecture, 
the formula for force and virial calculation, and the definition of the loss function.
We explained the RAW data format defined by DeePMD-kit, 
which provides a protocol for utilizing simulation data generated by other molecular simulation packages
and can be easily manipulated by text processing tools in the UNIX-like systems and Python.
We provided brief instructions on the model training, testing, and how to setup DeePMD simulation under LAMMPS and i-PI.
Finally
the accuracy and efficiency of the DeePMD-kit package is illustrated by an example of bulk liquid water system.

\newparagraph
The current version of DeePMD-kit only provides CPU implementation of the descriptor computation.
In the training stage this computation is  embarrassingly parallelized by OpenMP.
However, during the evaluation of energy, force, and virial in MD simulations, this computation is not parallelized.
In the future we will provide support on the parallel computation of descriptors via CPU multicore and GPU multithreading mechanisms.

\section{acknowledgments}
The work of H. Wang is supported by the National Science Foundation 
of China under Grants 11501039 and 91530322, 
the National Key Research and Development Program of China 
under Grants 2016YFB0201200 and 2016YFB0201203, 
and the Science Challenge Project No. JCKY2016212A502.
The work of L. Zhang, J. Han and W. E is supported in part by ONR grant N00014-13-1-0338, DOE grants DE-SC0008626 and DE-SC0009248, 
and NSFC grant U1430237.
Part of the computational resources is provided by the Special Program for Applied Research 
on Super Computation of the NSFC-Guangdong Joint Fund under Grant No.U1501501.


\appendix
\section{Deriviation of force and virial}
\label{app:tmp1}
By using Eq.~\eqref{eqn:step1} and \eqref{eqn:step2}, 
the force of the $i$-th atom is given by
\begin{align*}
  \vect {F}_i 
  =&\,
  -  \frac{\partial }{\partial \vect R_{i}} \sum_j E_{s(j)}\\
  =&\,
     -\sum_{j,k,\alpha} \frac{ \partial E_{s(j)}}{\partial D_{jk}^\alpha} \frac{\partial D_{jk}^\alpha(\vect R_{jk}, \vect R_{ja(j)}, \vect R_{jb(j)})}{\partial \vect R_{i}} \\
  =&
     -\sum_{k\in N(i),\alpha}\frac{ \partial E_{s(i)}}{\partial D_{ik}^\alpha} \frac{\partial D_{ik}^\alpha(\vect R_{ik}, \vect R_{ia(i)}, \vect R_{ib(i)})}{\partial \vect R_{i}} \\
   &
     -\sum_{j\neq i}\sum_{k\in N(j),\alpha} \delta_{i,a(j)}\frac{ \partial E_{s(j)}}{\partial D_{jk}^\alpha} \frac{\partial D_{jk}^\alpha(\vect R_{jk}, \vect R_{ja(j)}, \vect R_{jb(j)})}{\partial \vect R_{i}} \\
   &
     -\sum_{j\neq i}\sum_{k\in N(j),\alpha} \delta_{i,b(j)}\frac{ \partial E_{s(j)}}{\partial D_{jk}^\alpha} \frac{\partial D_{jk}^\alpha(\vect R_{jk}, \vect R_{ja(j)}, \vect R_{jb(j)})}{\partial \vect R_{i}} \\
   &
     -\sum_{j\neq i}\sum_{k\in \tilde N(j),\alpha} \delta_{i,k}\frac{ \partial E_{s(j)}}{\partial D_{jk}^\alpha} \frac{\partial D_{jk}^\alpha(\vect R_{jk}, \vect R_{ja(j)}, \vect R_{jb(j)})}{\partial \vect R_{i}}.
\end{align*}
The virial of the system is given by
\begin{align*}
  \Xi
  & = -\frac12 \sum_i \vect R_i \vect F_i \\
  & = \frac 12 \sum_i \vect R_i \sum_{j\neq i} \frac{\partial E}{\partial \vect R_{ij}}  
    + \frac 12 \sum_i \vect R_i \sum_{j\neq i} \frac{\partial E}{\partial \vect R_{ij}}  \\
  & = \frac 12 \sum_i \vect R_i \sum_{j\neq i} \frac{\partial E}{\partial \vect R_{ij}}  
    - \frac 12 \sum_i \vect R_i \sum_{j\neq i} \frac{\partial E}{\partial \vect R_{ji}}  \\
  & = \frac 12 \sum_i \vect R_i \sum_{j\neq i} \frac{\partial E}{\partial \vect R_{ij}}  
    - \frac 12 \sum_j \vect R_j \sum_{i\neq j} \frac{\partial E}{\partial \vect R_{ij}}  \\
  & = \frac12 \sum_{i\neq j} \vect R_{ij} \frac{\partial E}{\partial \vect R_{ij}}. \\
\end{align*}
By using Eq.~\eqref{eqn:step1} and \eqref{eqn:step2}, it reads
\begin{align*}
  \Xi
  =&\,
     \frac12 \sum_{i\neq j} \vect R_{ij} \frac{\partial }{\partial \vect R_{ij}} \sum_p E_{s(p)}\\
  =&\,
     \frac12 \sum_{i\neq j} \vect R_{ij}  \sum_{p,q,\alpha} \frac{ \partial E_{s(p)}}{\partial D_{pq}^\alpha} \frac{\partial D_{pq}^\alpha(\vect R_{pq}, \vect R_{pa(p)}, \vect R_{pb(p)})}{\partial \vect R_{ij}} \\
  =&\,
     \frac12 \sum_{i\neq j} \vect R_{ij} \sum_\alpha\frac{ \partial E_{s(i)}}{\partial D_{ij}^\alpha} \frac{\partial D_{ij}^\alpha(\vect R_{ij}, \vect R_{ia(i)}, \vect R_{ib(i)})}{\partial \vect R_{ij}}    \\
   &
     + \frac12 \sum_{i\neq j} \vect R_{ij} \delta_{j,a(i)} \sum_{q,\alpha} \frac{ \partial E_{s(i)}}{\partial D_{iq}^\alpha} \frac{\partial D_{iq}^\alpha(\vect R_{iq}, \vect R_{ia(i)}, \vect R_{ib(i)})}{\partial \vect R_{ij}} \\
   &
     + \frac12 \sum_{i\neq j} \vect R_{ij} \delta_{j,b(i)} \sum_{q,\alpha} \frac{ \partial E_{s(i)}}{\partial D_{iq}^\alpha} \frac{\partial D_{iq}^\alpha(\vect R_{iq}, \vect R_{ib(i)}, \vect R_{ib(i)})}{\partial \vect R_{ij}}. \\
\end{align*}

\end{document}